# Characterizing Neutral Modes of Fractional States in the Second Landau Level

M. Dolev, Y. Gross, R. Sabo, I. Gurman, M. Heiblum, V. Umansky and D. Mahalu
*Braun Center for Submicron Research, Dept. of Condensed Matter Physics,*
*Weizmann Institute of Science, Rehovot* 76100*, Israel*

**ABSTRACT**
**Quasiparticles, which obey non abelian statistics, were predicted to exist in different physical systems, but are yet to be observed directly. Possible candidate states, which are expected to support such quasiparticles, are the $v=8/3$, $v=5/2$ and $v=7/3$ fractional quantum Hall states (in the second Landau level). The non abelian quasiparticles are expected to carry charge and a unique form of a chiral neutral edge mode. Recent measurements in the $v=5/2$ state detected quasiparticle charge $e/4$ and an upstream (opposite to charge transport) chiral neutral mode; both agreeing with a non abelian anti-Pfaffian state; although not excluding the possibility of edge reconstruction as the source for the detected upstream neutral mode and a different type of state. Here we present results of detailed measurements of charge and neutral modes in the main three quantum Hall states of the second Landau level. For the $v=8/3$ state we found a quasiparticle charge $e/3$ and an upstream neutral mode - excluding the possibility of a non abelian Read-Rezayi state and supporting a Laughlin-like state. Such exclusion holds for the hole-conjugate $v=7/3$ state as well, in which no upstream neutral mode was detected. This also proves that edge reconstruction was not present in the $v=7/3$ state, suggesting its absence also in $v=5/2$ state and thus supporting further the non abelian anti Pfaffian state.**

The statistics of quantum particles determines the properties of their many body wavefunction under particles exchange. For ubiquitous particles with abelian statistics, an interchange of two identical particles adds a phase to the original two-particle wavefunction ($\pi$ for fermions, $2\pi$ for bosons, or $2\pi/m$ with $m$ integer, for anyons in two dimensions). Alternatively, for non abelian particles [1-3], with a highly degenerate ground state of the system, an interchange of two particles may shift the ground state of the system to an orthogonal (degenerate) ground state. Moreover, as their name suggests, different interchanges do not commute. Prime candidates for non abelian behavior are the charged excitations (quasiparticles) in specific fractional quantum Hall (FQH) states, which were not proven to exist yet.

The fractional quantum Hall effect (FQHE) is observed in two dimensional electron gas (2DEG), subjected to a strong perpendicular magnetic field. It is characterized by quantized plateaus in the Hall resistance, coinciding with zero longitudinal resistance. Three of the incompressible FQH states in the second Landau level (LL), which were predicted to support non abelian quasiparticles, are $v=5/2$, $v=8/3$ and its particle-hole conjugate $v=7/3$. Starting with $v=5/2$ state, it could, in principle, be described by a variety of wavefunctions. While some are abelian [4-6], others are non abelian [7-10], however, all are expected to support quasiparticles with $e/4$ charge - as was indeed measured recently [11-15]. Moreover, all the proposed states are expected to support chiral neutral edge modes, albeit of different nature. For example, the recent observation of an *upstream* neutral mode (with an opposite chirality to that of charge propagation) [16], most likely supports the non abelian anti-Pfaffian state [8, 9]; although edge reconstruction may lead to an upstream neutral mode [17] (of abelian or non abelian states). For the two conjugate states, $v=7/3$ and $v=8/3$, two main candidate states were proposed: an abelian, Laughlin-like state (similar in nature to the $v=1/3$ and $v=2/3$ states [18, 19]), and a non abelian Read-Rezayi state [20]. Note that although it is natural to assume that $v=7/3=2+1/3$ state is a Laughlin-like state, similar to the robust $v=1/3$, numerical calculations [21, 22] suggest that it might be described by a hole conjugate state of the non abelian k=4 Read-Rezayi state [20]. Such a non abelian variety of these states expected to support quasiparticle charge of $e/6$, with a chiral upstream neutral mode only in $v=7/3$ (the $v=8/3$ state should support a downstream neutral mode) [20]. Alternatively, the abelian version of $v=7/3$ and the $v=8/3$ states, is expected to mimic the $v=1/3$ and $v=2/3$ states, respectively - namely, charge of $e/3$ and no upstream neutral in $v=7/3$ case, and charge $e/3$ or $2e/3$ and upstream neutral mode in $v=8/3$ case [19].

Our measurements were performed on two similar samples, with a 2DEG embedded in a 30nm wide AlGaAs-GaAs-AlGaAs quantum well, doped on both sides, with an areal electron density $2.9 \times 10^{11} cm^{-2}$ and a low temperature mobility $29 \times 10^{6} cm^{2}/V$-s (both measured in the dark, Umansky *et al.* [23]). Hall measurement at 10mK (Fig. 1a), shows five significant fractional states, $v=11/5, 7/3, 5/2, 8/3$ & $14/5$ (with $R_{xx} \sim 0$ for $v=7/3, 5/2$ and $8/3$), measured on the ungated part of a Hall-bar-type sample (width $50 \mu m$ and total length $380 \mu m$, Fig. 1b). A single quantum point contact (QPC) constriction was formed by a negatively biased split gate on top of a narrower part of the mesa (width $5 \mu m$), with a distance of $40 \mu m$ between the closest ohmic contact and the QPC. Three types of measurements were performed: (a) Downstream noise measurements, in which current was driven through C1 or C2 and partitioned by the QPC constriction – leading to shot noise in M; (b) Upstream noise measurements, in which an *upstream* neutral mode emanates from N (in Fig. 1b). Impinging on the QPC constriction it is expected to generate shot noise [16] that is monitored at contact M (the *downstream* current is collected G1); (c) When an upstream neutral mode was found, its influence on charge partitioning in the QPC constriction (due to a simultaneously injected current from another contact) was measured.

For the analysis of the excess noise (the added noise when current was injected), we apply the 'single particle' model - used successfully before [11, 13, 24-29] - to determine particles' charge $e$, $e/3$, $e/5$, $e/7$, $e/4$ and $2e/3$ ($e/3$) at filling factors $v$=integer, 1/3, 2/5, 3/7, 5/2 and 2/3, respectively. The model is based on the assumption that quasiparticles are stochastically partitioned by a QPC constriction, and thus obey a binomial distribution with a variance proportional to the quasiparticle charge. Other models, based on backscattering of a chiral Luttinger liquid, were found to be inconsistent with the conductance and the excess noise produced in the QPC constriction (see [13] for more details). One has to be careful when calculating the excess noise for fractional states where multiple edge channels coexist (such as in the second LL). Partitioning the $i^{th}$ channel, flowing along the boundary between filling factors $v_i$ and $v_{i-1}$, leads to a low frequency spectral density of current fluctuations at finite temperature $T$ [30, 31]:

$$S^{i}(V_{sd})_T = 2e^{*}V_{sd}\Delta g_i\, t_{v_i-v_{i-1}}(1-t_{v_i-v_{i-1}}) \left[ coth(\frac{e^{*}V_{sd}}{2k_BT}) - \frac{2k_BT}{e^{*}V_{sd}} \right] + 4k_BTg \;, \qquad (1)$$

where $e^*$ is the partitioned quasiparticle charge, $V_{sd}$ is the applied DC excitation voltage, $\Delta g_i = g_i - g_{i-1}$, with $g_j = v_j e^2 / h$, and $t_{v_i - v_{i-1}} = \frac{g - g_{i-1}}{\Delta g_i}$ is the transmission probability of the $i^{\text{th}}$ channel, with $g$ the two-terminal (Hall) conductance. In our samples, for bulk filling factors $v_i = 7/3$, 5/2 and 8/3, the next lower lying state traversing freely the QPC constriction was $v_{i-1}=2$ [11]. Since the tunneling charge can be the fundamental (bulk) quasiparticle charge or its integer multiple (due to *bunching*) [19], its measurement sets an upper limit on its fundamental value [13]. This upper limit was obtained either when the differential transmission of the QPC constriction was energy independent [13]; at elevated electron temperature [13]; or in a simultaneous presence of a neutral mode in the QPC constriction [16].

We start with measurements performed at $v=8/3$. Injecting current through N and finding excess noise, with no average current, in M provides a direct proof for the existence of an upstream neutral chiral mode [16]. Noise detected in M can be attributed to the fact that an arrival of neutral mode at the QPC constriction induces tunnelling of charged quasiparticles, or equivalently, viewing the neutral mode as a stream of 'dipoles', which are being fragmented by the constriction, thus forming oppositely propagating particle-hole pairs. Alternatively, the heat transported by the upstream neutral mode may 'heat up' one side of the constriction, thus leading to an added noise [32]. While the amount of energy carried by the neutral mode is difficult to estimate, its upper bound can be estimated from the input power into N (typically, $V_N \sim 50\mu V$ & $I_N=5nA$, with the current flowing in the relevant 'upper 2/3 channel' $I_N^{8/3-2} = 1.25 nA$). Since the downstream charge mode carries power $\frac{1}{2} I_N^{8/3-2} V_N$, the other half, resulting usually with a hot spot on the opposite side of the contact N, is likely to be the source of the upstream neutral mode – providing thus an upper limit of energy carried by the neutral mode.

The excess noise in M, as function of the current $I_N^{8/3-2}$ in N, is shown in Fig. 2a for three different transmission probabilities $t_{8/3-2}$ of the QPC constriction (measured by driving a small AC signal ($2\mu V$ RMS at 910 kHz) into C1 and measuring the transmitted signal at M). The excess noise for $I_N^{8/3-2} = 1.25 nA$ is plotted for a few values of the average transmission probability in Fig. 2b. It roughly follows $t_{8/3-2}(1-t_{8/3-2})$, dropping to zero at $t_{8/3-2}=0$. This indicates that the two lower integer edge modes do not participate in the process of tunnelling. Moreover, and easier to measure, the effect of the upstream neutral mode on the transmission of the QPC constriction (measured for current injected from C1) was found to be significant only for rather low transmission probabilities (see Fig. 2c).

Observing an upstream neutral mode at $v=8/3$ is not consistent with a non abelian k=4 Read-Rezayi state [20]. Moreover, for a nearly energy independent $t_{8/3-2}$, with current injected from C1, the partitioned charge was determined to be $e/3$ (blue curves for transmission and noise in Figs. 3a & 3b) – again inconsistent with the predicted $e/6$ for the non abelian state, and thus pointing at a Laughlin '2/3-like' state. How will the partitioned quasiparticle charge be affected by the presence of a neutral mode? With $I_N=5nA$, the transmission and excess noise were measured versus the current injected at C1, as shown by the red curves in Figs. 3a & 3b, respectively. The striking effect is the increased temperature of the partitioned quasiparticles to 25mK, while quasiparticle charge remained $e/3$. At somewhat lower and energy dependent transmission (Fig. 3c), with an apparent partitioned charge *bunching* (~0.55*e*, [13]), the added upstream neutral mode affected the temperature, the transmission and the partitioned charge. The transmission turned to be energy independent (*linear*), the charge returned to nearly $e/3$ and the temperature raised to $T$=25mK.

Partitioning charge in the $v$=7/3 state led excess noise with a corresponding $e/3$ charge [13] for nearly energy independent transmission. This is in contradiction to the prediction of $e/6$ for an hole conjugate state of the k=4 Read-Rezayi state [20]. As seen in Fig. 4, an upstream neutral mode is absent - strengthening our conclusion that the $v$=7/3 state is a Laughlin-like state. Moreover, an edge reconstruction, which may have led to an upstream neutral mode, is not likely to take place. We compare the excess noise for $v$=7/3, $v$=5/2 and $v$=8/3 states due to fragmentation of the neutral mode as function $I_N^{v-2}$ (one can also plot this comparison as function of $V_N$) in Fig. 4. The fact that the excess noise for $v$=5/2 is smaller than for $v$=8/3 for similar transmission probabilities might indicate either that the carried energy by the mode is smaller or, alternatively, the decay length for the $v$=5/2 (as it traverses from N to the QPC constriction) is shorter [16].

We examined the most pronounced fractional states in the second LL via charge and neutral mode measurements, testing whether their nature might fit that of non abelian states. Our results for the $v$=8/3 and $v$=7/3 states suggest that they are Laughlin type states, with a fundamental charge of $e/3$ with an upstream neutral mode only in $v$=8/3. Hence, with its upstream neutral mode and charge of $e/4$, the $v$=5/2 state is the most likely candidate for a non abelian state, with an anti-Pfaffian wavefunction [8, 9, 16].

**Acknowledgements**
We thank Ed Rezayi, Bernd Rosenow, Alessandro Braggio, Roni Ilan, Yuval Gefen and Ady Stern for helpful discussions. We thank for the partial support of the Israeli Science Foundation (ISF), the Minerva foundation, the German Israeli Foundation (GIF), the German Israeli Project Cooperation (DIP), the European Research Council under the European Community's Seventh Framework Program (FP7/2007-2013) / ERC Grant agreement # 227716, and the US-Israel Bi-National Science Foundation (BSF).

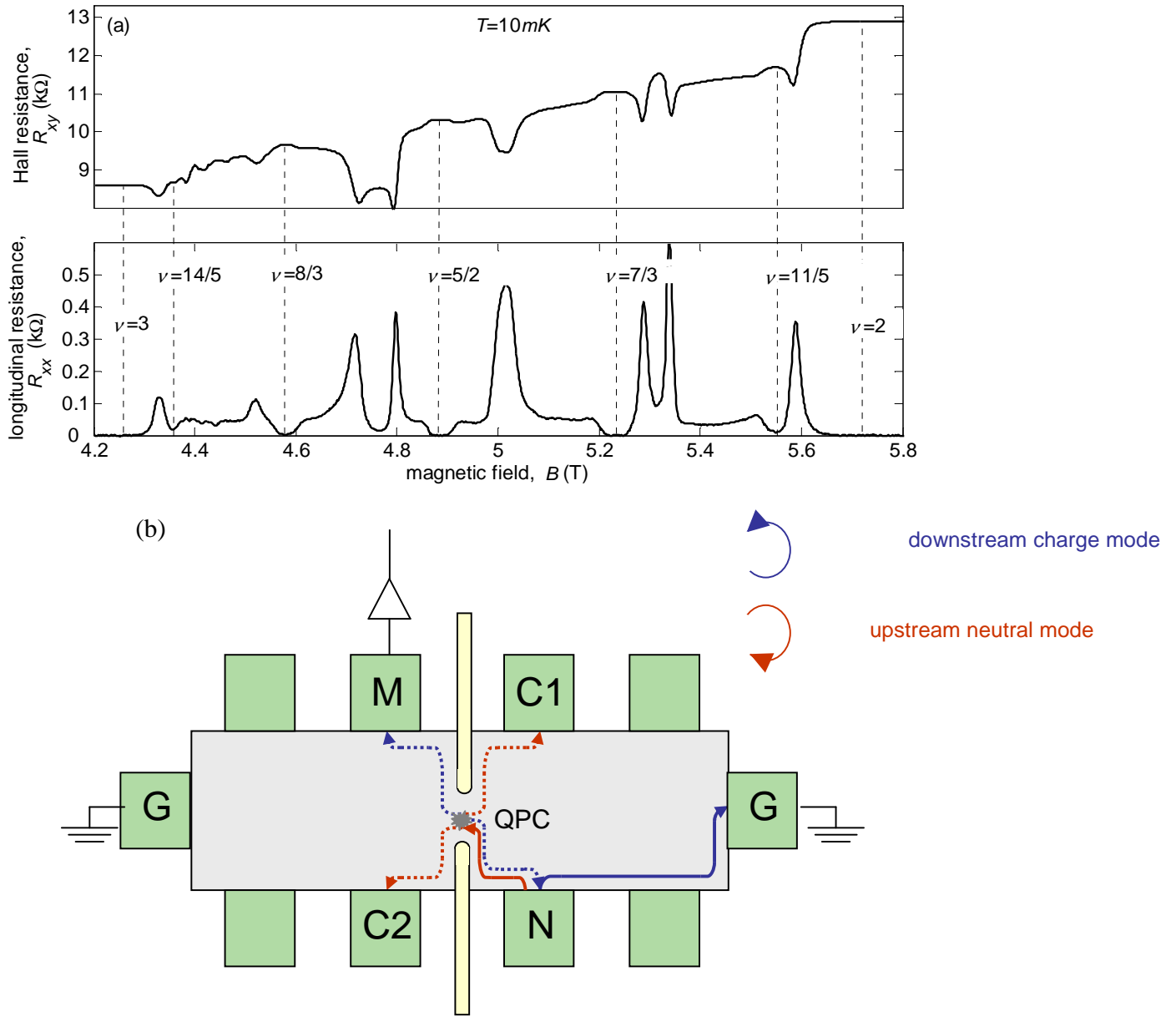

**Figure 1**
Hall effect data. (a) Fully developed fractional quantum Hall states are observed in $v=7/3$, $5/2$ and $8/3$ states. (b) Schematic description of the gated Hall bar. Chiral charge mode flows counter clockwise. Current driven into "charge-sources" C1 and C2 flows towards the QPC; it is being partitioned, and thus generating shot noise proportional to the tunneling charge, which is measured in probe M. Upstream neutral mode, if present, flows clockwise; presented by the red and blue trajectories that emerge from the "neutral source" N. The charge current (blue) flows into the ground G1. The upstream neutral mode (red) flows towards the QPC and generates charge mode fluctuations (dashed blue), which can be measured in M.

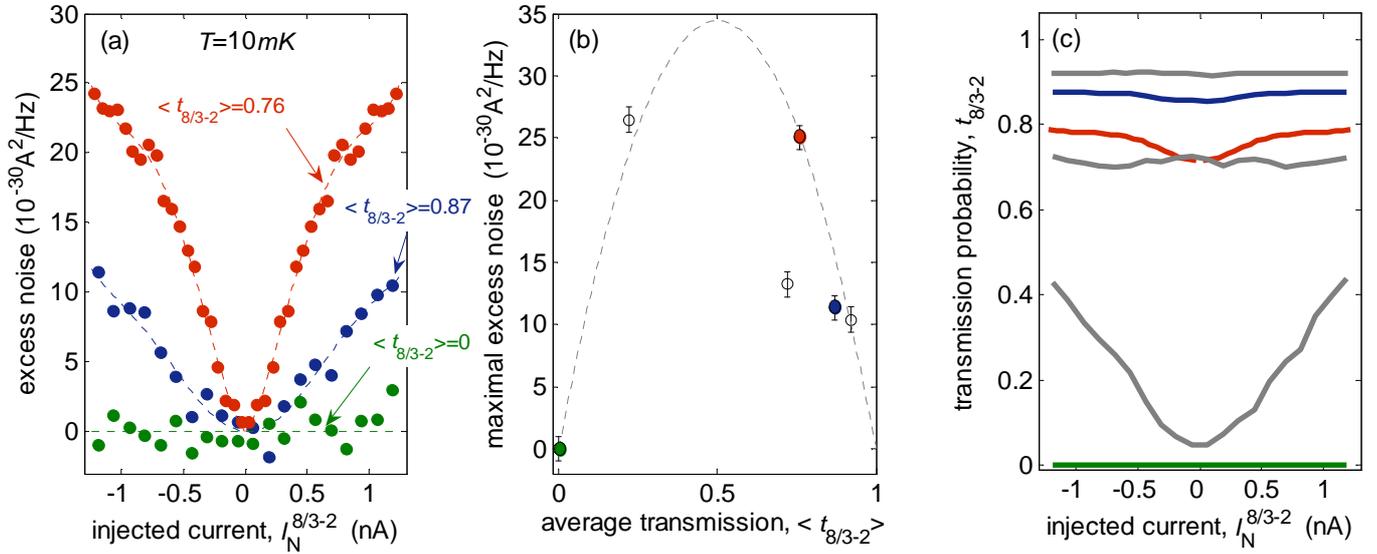

**Figure 2**

Excess noise and conductance in $v=8/3$ due to the arrival of a neutral mode at the QPC at $T=10mK$. (a) Excess noise as a function of the current in the upper 2/3 channel, $I_N^{8/3-2}$, injected from N (charge current flows into G1 and upstream neutral mode arrives at the QPC), for three different transmission probabilities of the QPC. Excess noise proves the existence of an upstream neutral mode. (b) Excess noise at $I_{8/3-2}=1.25nA$ as a function of the transmission probability of the QPC. Dashed line is proportional to $t_{8/3-2}(1-t_{8/3-2})$. (c) Differential transmission probability (measured by current in C1 and measurement at M) as a function of the current $I_N^{8/3-2}$ at N. Strong dependence of the transmission probability is observed at lower transmissions.

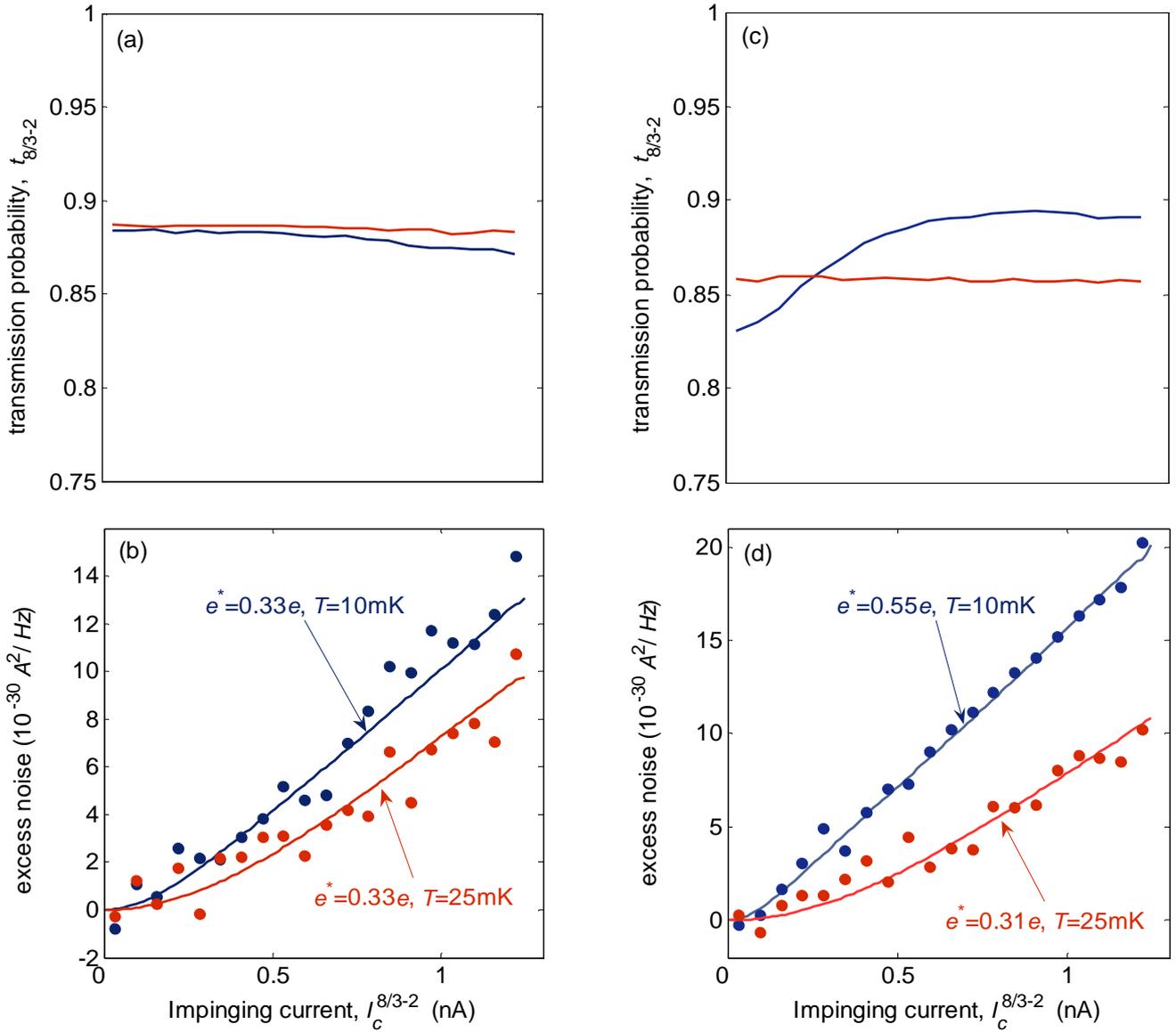

**Figure 3**

Transmission probability and shot noise in $v=8/3$ as a function of the impinging current (injected from C1), with (red) and without (blue) an additional upstream neutral mode (injected from N). Measured shot noise is in dots; solid lines are the expected shot noise for specific charges and temperatures, as indicated. (a-b) For transmission probability independent of the impinging current, charge equals to $e/3$ with and without the additional neutral current. (c-d) For transmission probability dependent on impinging current, the presence of neutral mode linearizes the transport as well as lowers the charge to almost $e/3$. In both cases the addition of the neutral mode raises the temperature to 25*mK*.

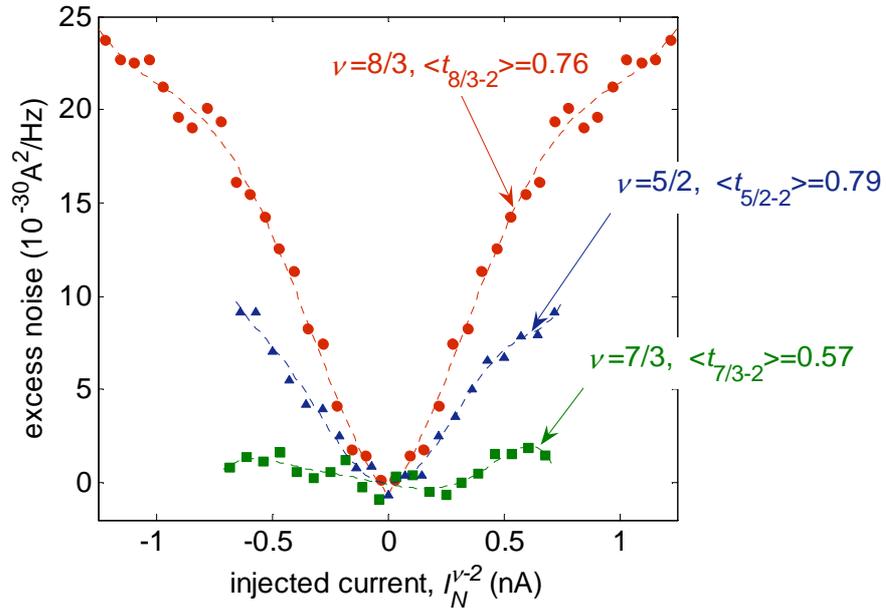

**Figure 4**
Comparison between the excess noise, as a function of the injected current from contact N, in $v=7/3, 5/2$ and $8/3$. Excess noise in $v=8/3$ and $5/2$ states proves the existence of an upstream neutral modes, oppositely to the case of $v=7/3$. Excess noise in the $v=8/3$ state was measured only for the positive injected current range, and was mirror imaged.